\documentclass[%
preprintnumbers,
nofootinbib,
 amsmath,amssymb,
 aps,
]{revtex4-2}

\usepackage{graphicx}
\usepackage{subcaption}
\usepackage[justification=raggedright]{caption}
\usepackage{dcolumn}
\usepackage{bm}
\usepackage{hyperref}
\hypersetup{
  colorlinks   = true, 
  urlcolor     = blue, 
  linkcolor    = blue, 
  citecolor   = red 
}

\usepackage{slashed}

\def\be{\begin{equation}}
\def\ee{\end{equation}}
\def\bea{\begin{eqnarray}}
\def\eea{\end{eqnarray}}
\def\bear{\begin{array}}
\def\ear{\end{array}}
\def\bfig{\begin{figure}}
\def\efig{\end{figure}}
\def\bcen{\begin{center}}
\def\ecen{\end{center}}
\def\bi{\begin{itemize}}
\def\ei{\end{itemize}}
\def\raw{\rightarrow}

\def\D{\displaystyle}

\def\d16{d_{16}}

\usepackage{bm}
\usepackage{xcolor}
\usepackage{multirow}
\usepackage{slashed}
\usepackage{physics}
\usepackage{makecell}

\newcommand{\no}{\nonumber}

\newcommand{\mL}{\mathcal{L}}

\newcommand{\mO}{\mathcal{O}}

\newcommand{\fpi}{F_\pi}
\newcommand{\mpi}{M_\pi}
\newcommand{\md}{\mathring{m}_\Delta}
\newcommand{\m}{\mathring{m}}
\newcommand{\g}{\mathring{g}_A}
\newcommand{\bfour}{\widetilde{b}_4}
\newcommand{\pthree}{\mO(p^3)}
\newcommand{\pfour}{\mO(p^4)}

\begin{document}

\title{Light-quark mass dependence of the nucleon axial charge\\ and pion-nucleon scattering phenomenology}

\author{Fernando Alvarado}
\email{Fernando.Alvarado@ific.uv.es}
\affiliation{Instituto de F\'isica Corpuscular (IFIC) and Departamento de F\'\i sica Te\'orica, \\ Consejo Superior de Investigaciones Cient\'{i}ficas (CSIC) and Universidad de Valencia (UV)\\ E-46980 Paterna, Valencia, Spain}
\author{Luis Alvarez-Ruso}%
\email{Luis.Alvarez@ific.uv.es}
\affiliation{Instituto de F\'isica Corpuscular (IFIC) and Departamento de F\'\i sica Te\'orica, \\ Consejo Superior de Investigaciones Cient\'{i}ficas (CSIC) and Universidad de Valencia (UV)\\ E-46980 Paterna, Valencia, Spain}

\date{December 28, 2021}

\begin{abstract}
The light-quark mass dependence of the nucleon axial isovector charge ($g_A$) has been studied up to next-to-next-to-leading order, $\pfour$, in relativistic chiral perturbation theory using extended-on-mass-shell renormalization, without and with explicit $\Delta (1232)$ degrees of freedom. We show that in the $\Delta$-less case, at this order, the flat trend of $g_A (\mpi)$ exhibited by state-of-the-art lattice QCD (LQCD) results cannot be reproduced using low energy constants (LECs) extracted from pion-nucleon elastic and inelastic scattering. A satisfactory description of these LQCD data is only achieved in the theory with $\Delta$. From this fit we report $g_A (M_{\pi\rm (phys)}) = 1.260 \pm 0.012$, close to the experimental result, and $d_{16}= -0.88\pm 0.88$~GeV$^{-2}$, in agreement with its empirical value. The large uncertainties are of theoretical origin, reflecting the difference between $\pthree$ and $\pfour$ that still persists at large $\mpi$ in presence of the $\Delta$.
\end{abstract}

\maketitle


\section{\label{sec:intro} Introduction}
The axial isovector charge $g_A^{u-d}$ ($g_A$ from now on) is a fundamental property of the nucleon related to the difference in the spin fractions carried by $u$ and $d$ quarks. With a magnitude precisely determined from neutron $\beta$ decay~\footnote{After taking into account radiative corrections, a numerical value of $g_{A} = 1.2754(13)_\mathrm{exp}(2)_\mathrm{RC}$ has been recently extracted~\cite{Gorchtein:2021fce} from the PDG average $g_A/g_V = 1.2756(13)$~\cite{Zyla:2020zbs}.}, the nucleon axial charge represents a benchmark for non-perturbative studies of Quantum Chromodynamics in the lattice (LQCD) alongside with other nucleon properties, such as scalar and tensor charges, electromagnetic form factors and parton distribution functions (see Ref.~\cite{Lin:2020reh} and Sec.~10 of Ref.~\cite{Aoki:2021kgd} for recent reviews). Over the last decade, progress in this direction has been significant~\cite{Aoki:2021kgd}, owing to better computational resources, improved algorithms and techniques to reduce systematic errors, in particular those induced by excited-state contamination, which can be considerable in the baryon sector for currently available lattice ensembles~\cite{Tiburzi:2009zp,Bar:2017kxh}.  

As the effective theory of QCD in the non-perturbative regime, chiral perturbation theory (ChPT) plays an important role. It allows to determine the light ($u$, $d$) quark mass dependence of low-energy hadronic observables in terms of low-energy constants (LECs) and perform model-independent extrapolations of LQCD results to the physical point. ChPT also allows to account for finite lattice-volume and lattice spacing corrections in a systematic way~\cite{Beane:2003xv,Beane:2004rf}. It has also proved helpful to deal with the contamination from excited states~\cite{Tiburzi:2009zp,Bar:2017kxh,RQCD:2019jai}. The interplay between ChPT and LQCD can also be used to determine poorly known LECs, which are difficult to access with experimental data. 
This is the case of $d_{16}$ present in the $\mathcal{O}(p^3)$ part of the $\pi N$ Lagrangian. Via a $4 d_{16} M_\pi^2$ term, this LEC controls the slope of the light-quark mass dependence of $g_A$.~\footnote{Let us recall that in the isospin-symmetric limit $m_u = m_d \equiv \hat{m}$, $\mpi^2 = 2 B_0 \hat{m} + \pfour$, so that, as customary, we indistinctly refer to the $\hat{m}$ or $\mpi$ dependence.} Therefore, its extraction from LQCD results at low pion masses is only natural.  LEC $d_{16}$ has been identified as one of the most significant sources of uncertainty in quark mass dependence of nuclear properties such as ground-state and binding energies through long-range nuclear forces~\cite{Beane:2002xf, Berengut:2013nh,Epelbaum:2013wla}.    

Alternatively, LECs have been extracted from experimental information~\footnote{Throughout this article LECs obtained from experimental data are called phenomenological or empirical to distinguish them from those extracted from LQCD simulations.} on various processes such as pion photo- and electroproduction~\cite{Bernard:1993bq,Hilt:2013fda,GuerreroNavarro:2020kwb,Rijneveen:2021bfw} but, predominantly, from pion-nucleon ($\pi N$) scattering~\cite{Mojzis:1997tu,Fettes:2000xg,Fettes:2000bb,Torikoshi:2002bt,Hoferichter:2015hva,Chen:2012nx,Alarcon:2012kn,Yao:2016vbz,Siemens:2016hdi}. Although $d_{16}$ cannot be extracted from $\pi N$ elastic scattering, it contributes significantly to inelastic, $\pi N \raw \pi \pi N$, scattering. Indeed, in a combined fit to both elastic and inelastic $\pi N$ scattering data  $d_{16}$ has been recently extracted, albeit with a large uncertainty~\cite{Siemens:2017opr}. 
Baryon ChPT (BChPT) has been used to calculate the axial charge within the heavy-baryon~\cite{Kambor:1998pi,Hemmert:2003cb,Bernard:2006te,Procura:2006gq,CalleCordon:2012xz} and relativistic~\cite{Schindler:2006it,Ando:2006xy,Procura:2006gq,Alarcon:2012kn,Chen:2012nx,Yao:2017fym,Lutz:2020dfi} ChPT approaches both without and with the $\Delta(1232)$ as an explicit degree of freedom. In the relativistic framework, both infrared and extended-on-mass-shell (EOMS) regularization methods have been applied. Some of these studies also address the nucleon axial form factor at low four-momentum transfers squared~\cite{Schindler:2006it,Ando:2006xy,Yao:2017fym,Lutz:2020dfi}.
The $\mpi$ dependence of $g_A$ has been specifically investigated in Refs.~\cite{Hemmert:2003cb,Bernard:2006te,Procura:2006gq,CalleCordon:2012xz,Yao:2017fym,Lutz:2020dfi}.


Armed with relativistic baryon ChPT, we revisit this problem up to next-to-next-to-leading order (NNLO) in the perturbative expansion. The EOMS scheme ensures that not only power counting but also analytic properties of loop functions are properly preserved. After describing in Sec.~\ref{sec:gAinBChPT} the details of the ChPT calculation and introducing the relevant terms of the effective Lagrangian we show in Sec.~\ref{sec:gApred} the $\mpi$ dependence of $g_A$ predicted with the phenomenological LECs obtained in Ref.~\cite{Siemens:2017opr} from $\pi N$ elastic and inelastic scattering. In order to obtain a better description  of recent LQCD data we have performed fits from which some of the LECs and, in particular, $\d16$, are determined. These are presented in Sec.~\ref{sec:gAfit}. A meaningful representation of the flat trend exhibited by the LQCD results is only achieved with explicit $\Delta(1232)$. Differences between orders, which are considerable between NLO and NNLO,  are used to provide a measure of the systematic error arising from the truncation of the perturbative series. These differences are the main source of uncertainty in the extracted $\d16$ value.

\section{\label{sec:gAinBChPT} The nucleon axial charge in relativistic BChPT}

In the isospin limit, the matrix element of the axial isovector current 
\bea
A_\mu^a(x)=\bar{q}(x) \gamma_\mu\gamma_5\frac{\tau^a}{2} q(x)  \,,
\eea
with $q=(u,d)^T$ the quark-field doublet, taken between on-shell nucleon states of equal four-momenta $p$ ($p^2 = m_N^2$), can be written as
\be
\langle N(p)|A_\mu^a(0)| N(p)\rangle =  \bar{u}(p) g_A \gamma_\mu \gamma_5 \frac{\tau^a}{\D 2} u(p) \,. \ee
The isovector character of the current is manifest given the presence of the Pauli isospin matrices $\tau^a$.

We calculate the axial charge $g_A$ employing relativistic BChPT with pions, nucleons and $\Delta$ as degrees of freedom. We use the standard power counting \cite{Weinberg:1991um}, extended to diagrams with $\Delta$ baryons following small scale expansion (SSE) of Refs.~\cite{Hemmert:1996xg,Hemmert:1997ye}. In this power counting, the mass difference $\delta = m_N-m_\Delta$ is $\mO(p)$.

\subsection{Relevant terms of the effective Lagrangian}
In this section the terms in the effective Lagrangian, $\mL_{\rm{eff}}$, 
required for our calculation are presented. We  need
\bea
\mL_{\rm eff} \supset \mL_{\pi N}^{(1)}+\mL_{\pi \Delta}^{(1)}+\mL_{\pi N \Delta}^{(1)}+\mL_{\pi N}^{(2)}+\mL_{\pi N \Delta}^{(2)}+\mL_{ \pi \Delta}^{(2)}+\mL_{\pi N}^{(3)}\ ,
\eea
where superscripts indicate the chiral order and subscripts show the present degrees of freedom. 
The terms in $\mL_{\pi N}^{(1,3)}$, $\mL_{\pi \Delta}^{(1)}$ and $\mL_{\pi N \Delta}^{(1)}$ that are relevant for our study have been displayed in Ref. \cite{Yao:2017fym} using the same notation adopted here. In addition,  $\mO(p^2)$ contributions are entailed to give $g_A$ at NNLO. 
Following the notation of Ref.~\cite{Fettes:2000gb}, the required terms of the $\pi N$ Lagrangian are
\begin{equation}
    \mL_{\pi N}^{(2)} \supset  \bar{\Psi}\left(c_1\langle \chi_+\rangle-\frac{c_2}{8 \m^2}\left(\langle u_\mu u_\nu\rangle \{D^\mu,D^\nu\}+\rm{h. c. }\right)+\frac{c_3}{2}\langle u_\mu u^\mu\rangle+\frac{i c_4}{4}[u_\mu,u_\nu]\sigma^{\mu\nu}\right)\Psi,
\end{equation}
where $\Psi$ is the isospin doublet of nucleon fields. Including only isovector axial external fields $a_\mu^a\tau^a/2  \equiv a_\mu$, we end up with
 \bea
 D_\mu&=&\partial_\mu+\Gamma_\mu\ ,\nonumber\\
  \Gamma_\mu&=&\frac{1}{2}[u^\dagger(\partial_\mu-i a_\mu)u+u(\partial_\mu+i a_\mu)u^\dagger]\ ,\nonumber\\
   u_\mu&=&i[u^\dagger(\partial_\mu-i a_\mu)u-u(\partial_\mu+i a_\mu)u^\dagger]\ ;
 \eea 
$\chi^+=u^\dagger\chi u^\dagger+u\chi^\dagger u$, with $\chi={\rm diag}(M_\pi^2,M_\pi^2)$, and $\langle ... \rangle $ denotes a trace over isospin.

In $\mL_{\pi N\Delta}^{(2)}$, after redundant terms are eliminated from  Eq.~(67) of Ref.~\cite{Jiang:2017yda} (see also Sec.~3.1 of Ref.~\cite{Holmberg:2018dtv} and the Appendix of Ref.~\cite{Unal:2021byi}) only the following monomials 

\be
\mL_{\pi N\Delta}^{(2)} \supset \bar{\Psi}_{\alpha}^k\xi^{\frac{3}{2}}_{ki}\left\{\frac{b_4}{2}\omega_\alpha^i\omega_\beta^j \gamma^\beta\gamma_5 \tau^j+\frac{b_5}{2}\omega_\alpha^j\omega_\beta^i \gamma^\beta\gamma_5 \tau^j \right\}\Psi+ \rm{h.c.}
\ee
contribute to the nucleon axial charge. Here $\Psi_\mu$ denotes the Rarita-Schwinger field of the $\Delta$ resonance. States $\{ \xi^{\frac{3}{2}}_{ij} \Psi_\mu^j \}_{i=1-3}$, where $\xi^{\frac{3}{2}}_{ij} = \delta_{ij} - \tau_i \tau_j /3  $ are isospin-$3/2$ projectors, are also isospin doublets, whose explicit expressions in terms of the physical $\Delta$ states are derived, for example, in Appendix~A of Ref.~\cite{Hacker:2005fh}. The isovector projection of $u^\mu$ is represented by $\omega^{\mu,i}=\frac{1}{2}\,\langle\tau^i u^\mu \rangle$ (see for instance Ref.~\cite{Jiang:2017yda}). Finally, also at $\mO(p^2)$~\cite{Jiang:2017yda,Siemens:2020vop}, 
\begin{equation}
\mL_{\pi\Delta}^{(2)}  \supset \bar{\Psi}_{\mu}^i\xi^{\frac{3}{2}}_{ij}\left\{a_1 \langle\chi_+\rangle \delta^{jk}g_{\mu\nu}\right\}\xi^{\frac{3}{2}}_{kl}\Psi^{l\nu}.
\end{equation} 
introduces an $\mpi$ dependent correction to the $\Delta$ mass, in the same way as the term proportional to $c_1$ does for the nucleon mass. 

\subsection{\label{subsec:gATheo} Perturbative calculation}

\begin{figure}[h]
\begin{subfigure}[t]{0.12\textwidth}
    \centering\includegraphics[width=\textwidth]{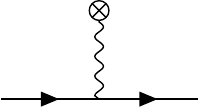}
    \caption{}
  \end{subfigure}\hfill
  \begin{subfigure}[t]{0.12\textwidth}
    \centering\includegraphics[width=\textwidth]{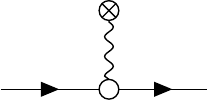}
   \caption{}
  \end{subfigure}\hfill
  \begin{subfigure}[t]{0.12\textwidth}
    \centering\includegraphics[width=\textwidth]{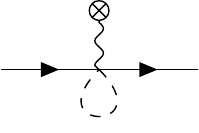}
   \caption{}
  \end{subfigure}\hfill
  \begin{subfigure}[t]{0.12\textwidth}
    \centering\includegraphics[width=\textwidth]{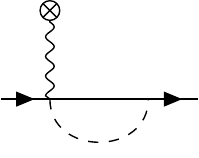}
    \caption{}
  \end{subfigure}\hfill
  \begin{subfigure}[t]{0.12\textwidth}
    \centering\includegraphics[width=\textwidth]{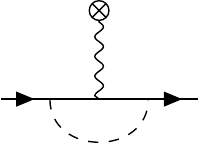}
    \caption{}
  \end{subfigure}\hfill
  \begin{subfigure}[t]{0.12\textwidth}
    \centering\includegraphics[width=\textwidth]{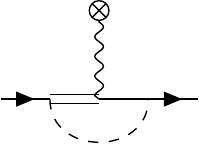}
    \caption{}
  \end{subfigure}\hfill
  \begin{subfigure}[t]{0.12\textwidth}
    \centering\includegraphics[width=\textwidth]{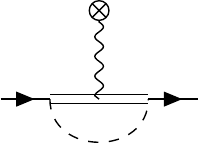}
    \caption{}
  \end{subfigure}
  
  \caption{Diagrams at orders $\mO(p)$, (a), and $\pthree$, (b)-(g), contributing to the nucleon axial charge. Dashed, solid single and double lines denote pions, nucleons and $\Delta$ resonances in that order; wiggly lines stand for external axial fields. The open circle represents an  $\pthree$ vertex, while the rest of the vertices are $\mO(p)$. Permutations of diagrams (d) and (f) have been omitted in the figure.}
\label{fig:p3diagrams}
\end{figure}

\begin{figure}[h]
\begin{subfigure}[t]{0.12\textwidth}
\addtocounter{subfigure}{7}
    \centering\includegraphics[width=\textwidth]{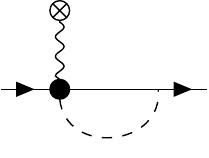}
    \caption{}
  \end{subfigure}\hfill
  \begin{subfigure}[t]{0.12\textwidth}
    \centering\includegraphics[width=\textwidth]{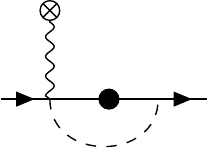}
    \caption{}
  \end{subfigure}\hfill
  \begin{subfigure}[t]{0.12\textwidth}
    \centering\includegraphics[width=\textwidth]{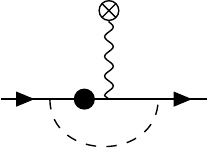}
    \caption{}
  \end{subfigure}\hfill
  \begin{subfigure}[t]{0.12\textwidth}
    \centering\includegraphics[width=\textwidth]{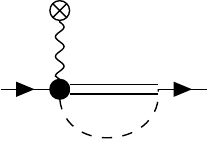}
    \caption{}
  \end{subfigure}\hfill
  \begin{subfigure}[t]{0.12\textwidth}
    \centering\includegraphics[width=\textwidth]{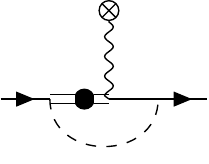}
    \caption{}
  \end{subfigure}\hfill
  \begin{subfigure}[t]{0.12\textwidth}
    \centering\includegraphics[width=\textwidth]{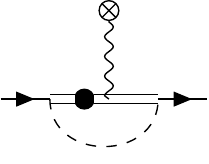}
    \caption{}
  \end{subfigure}\hfill
  \begin{subfigure}[t]{0.12\textwidth}
    \centering\includegraphics[width=\textwidth]{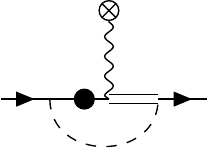}
    \caption{}
  \end{subfigure}
  \caption{$\pfour$ diagrams contributing to the nucleon axial charge. Line styles are the same as in Fig.~\ref{fig:p3diagrams}. Filled circles denote  $\mO(p^2)$ vertices. Permutations of all these diagrams are taken into account but not explicitly represented. 
}
\label{fig:p4diagrams}
\end{figure}

The set of Feynman diagrams that contribute to the axial form factor and, in particular, to $g_A$ are displayed in Fig.~\ref{fig:p3diagrams} and \ref{fig:p4diagrams}. The LECs that each of these diagrams brings about are listed in Table \ref{tab:diagLECs}.
At $\mO(p^4)$, there are contributions from $\mO(p^2)$ vertices but also baryon ($N, \Delta$) mass insertions. The later are calculated perturbatively, i.e. we evaluate directly diagrams (i), (j), (l)-(n) of Fig.~\ref{fig:p4diagrams}, avoiding Dyson re-summations to all orders in the propagators.  Alternative choices have been considered in Refs.~\cite{Ando:2006xy,Lutz:2020dfi}. 

\begin{table}[h!]
\caption{LECs introduced by Feynman diagrams in  Figs. \ref{fig:p3diagrams},  \ref{fig:p4diagrams} and by wave-function renormalization (wfr).}
\centering
  \begin{tabular}{|c|c|c|c|}
  \hline
      Diagrams & $\mO(p)$ & $\mO(p^2)$ & $\pthree$ \\
    \hline
    (a), (c), (d) & $\g$ & - & - \\
    (b) & - & - & $d_{16}$ \\
    (e) & $\g$ & - & - \\
    (f) & $\g$, $h_A$ & - & - \\
    (g) & $g_1$, $h_A$ & - & - \\
    (h) & $\g$ & $c_{2-4}$ & - \\
    (i) & $\g$ & $c_{1}$ & - \\
    (j) & $\g$ & $c_{1}$ & - \\
    (k) & $h_A$ & $b_{4,5}$ & - \\
    (l) & $\g$, $h_A$ & $a_1$ & - \\
    (m) & $g_1$, $h_A$ & $a_1$ & - \\
    (n) & $\g$, $h_A$ & $c_1$ & - \\
    wfr & $\g$, $h_A$  & $c_1$, $a_1$ & - \\
    \hline
  \end{tabular}
  \label{tab:diagLECs}
\end{table}

Although not diagrammatically represented, nucleon wave-function renormalization is taken into account in the standard way. To the order of our calculation, only the $\mO(p)$ diagram (a), giving just the axial charge in the chiral limit $\g$, should be multiplied by the wave-function renormalization constant $Z_N$, calculated from the $\mO(p^4)$ nucleon self-energy. 
\be
\label{eq:wfr1}
Z_N \approx 1 + \left. \frac{\partial \Sigma_N^{(4)}}{\partial \slashed{p}}\right|_{\slashed{p}=m_N} \,.
\ee
Additional  $\mO(p^3)$ and $\mO(p^4)$ contributions to $g_A$ are generated in this way. The LECs that enter in them are also listed in Table~\ref{tab:diagLECs}. 

In order to absorb the ultraviolet divergencies generated by loops, we perform dimensional renormalization in the $\widetilde{\rm{MS}}$ scheme~\cite{Fuchs:2003qc}, 
at the scale $\mu=\m$, where $\m$ denotes the nucleon mass in the chiral limit.  An additional renormalization is then performed to account for the power-counting breaking caused by the presence of baryon masses which do not vanish in the chiral limit. Among the available schemes we employ EOMS renormalization~\cite{Fuchs:2003qc}  which consists in the absorption of the power counting breaking (PCB) terms in a redefinition of the LECs. In this way, power counting is restored without altering the analytic properties of the loops and preserving covariance. Note that, as in earlier studies~\cite{Yao:2016vbz,Yao:2017fym}, PCB terms are identified and subtracted in an expansion in powers or $\mpi$ but not in $\delta$. The EOMS renormalization shifts for the LECs are lengthy, so we have included them in the supplementary material.

In the way outlined above we obtain the axial charge within the EOMS renormalization scheme up to $\pfour$ with explicit $\Delta$. Our result has the following structure, with superindices indicating the chiral order:
\bea \label{eq:gAexp}
    g_A &=& 
    \g+4 d_{16}\mpi^2+g_{A\rm{(loop)}}^{(3)\slashed{\Delta}}(\g;\mpi)+g_{A\rm{(loop)}}^{(3)\Delta}(\g,h_A,g_1;\mpi)\no\\
    && +g_{A\rm{(loop)}}^{(4)\slashed{\Delta}}(\g,c_1,c_2,c_3,c_4;\mpi)+g_{A\rm{(loop)}}^{(4)\Delta}(\g,h_A,g_1,c_1,a_1,b_4,b_5;\mpi)+\mO(p^5)\,.
\eea
The structures arising from the loops are preserved at the price of keeping higher order terms. The $\pthree$ part of $g_A$ is given in  Eq.~(A4) of Ref.~\cite{Yao:2017fym}. In Appendix~\ref{app3} we provide the $\pfour$ contribution from wave-function renormalization. The length of the rest of the $\pfour$ expression, which depends on several LECs, discourages us from giving it explicitly but is available in a Mathematica notebook as supplementary material.  
We nonetheless identify that at $\pfour$, i. e. at $\mO(\mpi^3)$ for $g_A$, $c_3$ and $c_4$ enter in the following combination $\widetilde{c}_4=c_4-c_3/2$, while $c_2$ enters only at $\mO(p^5)$ [$\mO(\mpi^4)$ in $g_A$].  
Although the rather involved contribution of diagram (k) does not, a priori, show the factorization of any mixture of $b_{4}$ and $b_5$ LECs, after expanding in $\mpi$, one finds that the combination which actually enters at $\pfour$, is $\bfour=b_4+ (12/13)\, b_5$.
  At $\mO(p^3)$ we reproduce the results of Eq.~(A4) of Ref.~\cite{Yao:2017fym}. At $\pfour$ we agree with the $\slashed{\Delta}$ EOMS expressions given in Ref.~\cite{Ando:2006xy,Chen:2012nx} except for the different treatment of nucleon mass insertions, which in our case, do not include resummations. As a result, $c_1 M_\pi^2$ factors appear only linearly in our calculation but not at all orders. Furthermore, a PCB term proportional to $c_4$ present in Ref.~\cite{Chen:2012nx} after the EOMS renormalization is absent in our final result.


\section{\label{sec:gApred} Light-quark mass dependence of the axial coupling from phenomenological input}

Once the framework and the chiral order are established, the $\mpi$ dependence of $g_A$ is ultimately determined by the LEC values. In Ref.~\cite{Siemens:2017opr}, elastic $\pi N$ and inelastic  $\pi N \raw \pi \pi N$ scattering has been studied up to $\mO(p^4)$ in covariant ChPT using a modified version of the EOMS approach~\cite{Siemens:2016hdi}. The LECs that enter $g_A(\mpi)$ at $\mO(p^4)$ in the $\Delta$-less model were extracted, including $d_{16}$, owing to the inclusion of low energy total and double differential $\pi N \raw \pi \pi N$ cross section data in the combined analysis. To make a prediction of $g_A(\mpi)$ based on this phenomenological input, we should transform the LECs from the modified EOMS of Refs.~\cite{Siemens:2016hdi, Siemens:2017opr}  to the conventional one adopted here. To the order we are working at, this transformation, whose details are disclosed in Appendix~\ref{app2}, affects the numerical value of $d_{16}$ but not of $c_{1-4}$. The axial coupling in the chiral limit, $\g$, which is not extracted in Refs.~\cite{Siemens:2016hdi, Siemens:2017opr}, is determined as 
\begin{equation}
\label{eq:gAphys}
g_A(\mpi = M_{\pi \rm (phys)},\g,d_{16},c_i)=g_{A\rm (phys)}=1.2754(13)_\mathrm{exp}(2)_\mathrm{RC}  
\end{equation}
from the experimental value, precisely known from $\beta$ decay~\cite{Gorchtein:2021fce}. Up to higher orders for $g_A$, the remaining parameters, i.e. the pion decay constant and the nucleon mass in the chiral limit, have been fixed to  $F_\pi\simeq F_{\pi\rm (phys)}=92.2$~MeV and $\m\simeq m_{N\rm (phys)}+ 4 c_1  M_{\pi\rm (phys)}^2$, with $M_{\pi\rm (phys)}=135$~MeV and $m_{N\rm (phys)}=939$~MeV, respectively. The numerical values for $d_{16}$, $c_{1-4}$ as well as $\g$ and their uncertainties are summarized in Table~\ref{tab:gApredLECs}.     
\begin{table}[h!]
\caption{LECs  used to obtain the $g_A(\mpi)$ dependence in the $\slashed{\Delta}$ case.}
\centering
  \begin{tabular}{|c|c|c|}
  \hline
      & $\pthree$ & $\pfour$ \\
    \hline
    $\g$  & $1.251\pm 0.051 $ & $1.089\pm 0.030$ \\
    $d_{16}$ (GeV$^{-2}$) & $-2.2\pm 1.1$  & $-1.86\pm 0.80$ \\
    $c_1$ (GeV$^{-1}$) & - & $-0.89\pm 0.06$  \\
    $c_2$ (GeV$^{-1}$) & - & $3.38\pm 0.15$  \\
    $c_3$ (GeV$^{-1}$) & - & $-4.59\pm 0.09$  \\
    $c_4$ (GeV$^{-1}$) & - & $3.31\pm 0.13$    \\
    \hline
  \end{tabular}
  \label{tab:gApredLECs}
\end{table}

\par
\begin{figure}[h!]
\centering
\includegraphics[scale=1.2]{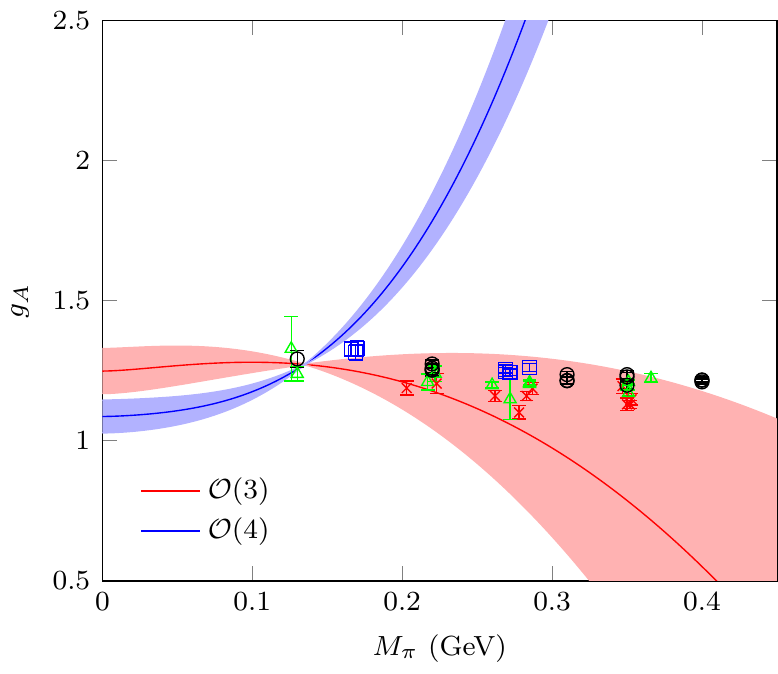}
\caption{Pion-mass dependence of $g_A$ at $\mO(p^3)$ (red) and $\mO(p^4)$ (blue) using phenomenological input from Ref.~\cite{Siemens:2017opr} and 1$\sigma$ error bands. The LQCD data points are from CalLat~18~\cite{Chang:2018uxx} (black circles), Mainz~19~\cite{Harris:2019bih} (red crosses), RQCD~19~\cite{RQCD:2019jai} (green triangles) and NME~21~\cite{Park:2021ypf} (blue squares).}
\label{fig:gApred}
\end{figure}

 
The resulting $g_A(\mpi)$ at $\mO(p^3)$ and $\mO(p^4)$ are displayed in Fig.~\ref{fig:gApred} together with a subset of recent LQCD determinations~\footnote{The selection criteria for the LQCD data are discussed in \ref{subsec:data}}. The curves are accompanied by 1$\sigma$ statistical error bands arising from the uncertainties in the LECs of Table~\ref{tab:gApredLECs} assuming they are Gaussian-distributed. Unaccounted correlations are negligible because the uncertainty is strongly dominated by the $d_{16}$ error. 
The $\mO(p^3)$ result shows agreement with the lattice determinations, albeit with increasing tension as $\mpi$ grows and a curvature in the central value that is absent in the data. On the other hand, it is immediately apparent that the $\mO(p^4)$ prediction not only does not improve the $\mO(p^3)$ but plainly fails to describe the $\mpi$ dependence of LQCD data. We have checked that alternative $c_{1-4}$ determinations from earlier $\pi N$ analyses using EOMS~\cite{Chen:2012nx,Yao:2016vbz} do not mitigate the steep rise of $g_A(\mpi)$ at $\mO(p^4)$. Therefore, the inability to reconcile the light-quark mass dependence of $\g$ at $\mO(p^4)$ with phenomenology, earlier observed in non-relativistic heavy-baryon ChPT~\cite{Bernard:2006te,Procura:2006gq}, is also a feature of the relativistic version of the theory.

The problem with the $\mO(p^4)$ result might be caused by an accidental slow convergence of $g_A$ in BChPT. Instead, one could argue that additional degrees of freedom such as the $\Delta$~\cite{Procura:2006gq} or even the Roper $N(1440)$ resonance might solve the issue. In Ref. \cite{Siemens:2017opr}, a version of the theory incorporating the $\Delta$ pole is also considered. However, in order to predict $g_A(\mpi)$ up to $\pfour$ with explicit $\Delta$, additional LECs absent in that study are needed. For this reason we do not tackle the role of the $\Delta$ in the $g_A(\mpi)$ dependence from purely empirical input but postpone it to the next Sec.~\ref{sec:gAfit} where our fits to recent LQCD data are reported. 


\section{\label{sec:gAfit} Analysis of LQCD results and LEC determination}


\subsection{\label{subsec:data} Data set and fit strategy}
The axial charge has been historically underpredicted in the lattice (see for instance Fig.~2 of Ref.~\cite{Constantinou:2017lok}), but, as a result of conceptual and technical improvements, the majority of LQCD results nowadays agree with the experimental value at the level of a few percent~\cite{Aoki:2021kgd}. Most significantly, the analysis of excited-state contamination has notably evolved in last years. Therefore, we only include in our data set results from studies with an improved treatment of these effects. We take into account renormalized $\{ g^i_A\}$ data at different $\{\mpi^i,a_i\}$ values, where $a$ stands for the lattice spacing, from~\footnote{The names are borrowed from the FLAG Review 2021~\cite{Aoki:2021kgd}.}  CalLat~18~\cite{Chang:2018uxx}, Mainz~19~\cite{Harris:2019bih}, RQCD~19~\cite{RQCD:2019jai}~\footnote{We only consider the simulations with $m_s\sim m_{s\rm{(phys)}}$, the ones suitable for an SU(2) ChPT analysis.} and NME~21~\cite{Park:2021ypf}~\footnote{We take the results from the fit strategy labelled as $\{4^{N \pi}, 3^*\}$, used to control excited-state contamination, averaging over the two renormarization methods $Z_1$ and $Z_2$.}. 
Our analysis treats 2+1+1 (CalLat~18) and 2+1 (Mainz~19, RQCD~19, NME~21) ensembles on the same footing, assuming that the $c$-quark sea content plays a negligible role. 
We disregard LQCD determinations of $g_A$ coming from  $q^2$, finite volume, $a$ or $\mpi$ extrapolations. Since we do not correct the finite volume effects, we consider only large volumes, satisfying $\mpi L\geq 3.5$, so that the neglected extrapolation is small and can be absorbed in the errors.

\par 
In order to gauge the ability of BChPT at $\pthree$ and $\pfour$ to describe the $\mpi$ dependence of $g_A$, and to extract $\g$ and $d_{16}$ LECs, we perform fits to the LQCD data set defined above. 
With this aim we define the following $\chi^2$:
\be \label{eq:chi2}
\chi^2=\sum_i \frac{\left( g_A (M^i_\pi,a_i)- g^i_A \right)^2}{(\Delta g^i_A)^2} + \chi^2_{\rm prior} \,.
\ee
In addition to $g_A(\mpi)$ from BChPT, our theoretical parametrization incorporates lattice discretization corrections as
\be
g_A(\mpi^i,a_i)=g_A(\mpi^i)+ x_j a_i^{n_j} \,.
\ee
Free parameters $x_j$, with $j=1,2,3,4$ when lattice data point $i \in \{\text{CalLat~18, Mainz~19, RQCD~19, NME~21}\}$ respectively, control the leading $a$-dependence of the  LQCD data, which is action specific: $n_{1,4} = 1$, while $n_{2,3} =2$.
These discretization corrections are small and do not substantially change the extracted LECs, but result in an appreciable reduction of the fits' $\chi^2/\rm{dof}$. 

Some of the LECs upon which $g_A(\mpi)$ depends are well determined in other studies and are kept fixed, while others are treated as free parameters together with $\g$ and $d_{16}$. To improve our description of the LQCD data and reduce correlations~\cite{Wesolowski:2015fqa}, for free LECs we assume naturalness $\Lambda^{n-1} \, c_n  \sim 1$, encoded in Gaussian priors
\begin{equation}
    \chi^2_{\rm prior} = \sum_{\text{free LECs}} \left(\frac{\Lambda^{n-1} \, c_n}{5} \right)^2 \,.
\end{equation}
Here, $c_n$ generically denotes a LEC of chiral order $\mO(p^n)$; the breakdown scale is set to $\Lambda = 1$~GeV$\sim \, 4 \pi F_\pi$~\cite{Manohar:1983md,Scherer:2012xha}.  We anticipate that a prior for $\g$ is superfluous, since its value is always driven  to a natural one by low $\mpi$ $g^i_{A}$ LQCD data. 

The large relative contribution of $\pfour$ terms discussed in the previous section and illustrated by Fig.~\ref{fig:gApred} is a clear indication that the theoretical error associated with truncation of the perturbative expansion should be taken into account. We follow the method proposed in Ref. \cite{Epelbaum:2014efa}. Let $X$ be an observable with a chiral perturbative expansion
\be\label{eq:X}
X = X^{(0)} + \sum_{m=1}^\infty \Delta X^{(m)} \,;
\ee 
$\Delta X^{(m)} = X^{(m)} - X^{(m-1)}$ encompasses all the monomials that start at order $m$. If $X$ is calculated up to order $n$, $X \approx X^{(n)}$, assuming that the truncation error is dominated by order $n+1$, its contribution $\Delta X^{(n+1)}$ can be estimated in a conservative way as~\cite{Epelbaum:2014efa,Siemens:2016hdi}
\begin{equation}
\label{eq:DX}
    |\Delta X^{(n+1)}| = \max\left\{Q^{n+1}|X^{(0)}|, Q^n|\Delta X^{(1)}|,..., Q |\Delta X^{(n)}| \right\} \,,
\end{equation}
where $Q$ denotes the expansion variable.

For $g_A$, $Q=\mpi/\Lambda$. Recalling that in this case $\mO(p^{5}) = \mO(\mpi^4)$, for our $\pfour$ calculation,  Eq.~(\ref{eq:DX}) gives   
\bea \label{eq:DgA4}
\Delta g_{A\chi}^{(5)}=&\max &\left\{\left(\frac{\mpi}{\Lambda}\right)^4\left|\g\right|, \left(\frac{\mpi}{\Lambda}\right)^2\left|\Delta g_A^{(3)}\right|,\frac{\mpi}{\Lambda} |\Delta g_A^{(4)}|\right\}\sim Q^4 = \mO(p^{5}) \,.
\eea
Connecting to Eq.~(\ref{eq:gAexp}), $\Delta g_A^{(3)} = 4 d_{16} \mpi^2 + g_{A \rm{(loop)}}^{(3)}$ and $\Delta g_A^{(4)} = g_{A \rm{(loop)}}^{(4)}$. Based on the results in Sec.~\ref{sec:gApred} it is easy to foresee that at larger $\mpi$, $\Delta g_{A\chi}^{(5)}$ will be determined by the last term in Eq.~(\ref{eq:DgA4}). In our $\pthree$ fits we do not assume any prior knowledge about $\pfour$ and, therefore, the truncation error is given by
\bea \label{eq:DgA3}
\Delta g_{A\chi}^{(4)}=&\max &\left\{\left(\frac{\mpi}{\Lambda}\right)^3\left|\g\right|, \frac{\mpi}{\Lambda}\left|\Delta g_A^{(3)}\right|\right\}\sim Q^3 = \mO(p^{4}) \,.
\eea
In our $\chi^2$, Eq.~\eqref{eq:chi2}, this theoretical error is added in quadrature to the one of LQCD points: 
\be
\label{eq:toterror}
(\Delta g^i_A)^2  =  (\Delta g^i_{A \rm{LQCD}})^2 + (\Delta g_{A\chi}(\mpi^i))^2 \,. 
\ee
This recipe assigns larger uncertainties to points at high $\mpi$, where the convergence of the chiral expansion is poorer, therefore reducing their impact on the fit. We should also mention that, as discussed in Refs.~\cite{Epelbaum:2014efa,Siemens:2016hdi,Furnstahl:2015rha}, although this theoretical error estimate does not have a clear statistical interpretation, results are similar to those obtained in a preferable but beyond the scope of the present study Bayesian approach~\cite{Wesolowski:2015fqa}. 

Fits are carried out iteratively until convergence is achieved. The first minimization is performed without theoretical errors, which are subsequently evaluated using the LECs determined in the previous step.  The lattice discretization parameters $x_j$ are fixed in the first iteration because evolving them results in overfitting. With the described procedure, LQCD and truncation errors are not independent and it is not obvious how to combine them in the error for a given observable. Consequently, following Ref.~\cite{Siemens:2016hdi}, we plot them separately in the error bands for $g_A (\mpi)$. Furthermore, as a quantitative measure of the agreement of our best-fit curve with the LQCD data, we also provide the $\chi^2_0$ value, defined as
\begin{equation}
\label{eq:chi20}
    \chi^2_0= \sum_i \frac{\left( g_A (M^i_\pi,a_i)- g^i_A \right)^2}{(\Delta g^i_{A \rm{LQCD}})^2} \,.
\end{equation}

\par 
Finally, we have tested the convergence range of our calculation by varying the maximum $\mpi$ of the lattice data included in the fits in the range of $M_{\pi \rm cut} \in [200, 402]$ MeV. A plateau in the $\chi^2$ and the fitted LECs is found towards the end of the interval. In consequence we report results including all points in the chosen data sets up to $\mpi = 402$~MeV~\footnote{This upper limit allows to include available points close but above $\mpi=400$ MeV.}. As discussed in Sec.~\ref{subsubsec:p4Dfull}, the theoretical error becomes larger at higher $\mpi$ where the convergence is poorer and, therefore, the corresponding LQCD points weight less in the fits. 

\subsection{\label{subsec:withoutD} $\slashed{\Delta}$ case}
\subsubsection{$\pthree$}

At $\pthree$ the only free parameters (besides the $x_j$ governing the $a$-dependence) are $\g$ and $d_{16}$ LECs. The results of the fit are shown in the upper left panel of Fig. \ref{fig:gAfit} and the first column of Table \ref{tab:gAfitLECs}. At the first sight, this model describes LQCD data well, with a good $\chi^2/\rm{dof}$, relatively small errors and natural $\g$ and $d_{16}$. As apparent from the comparison of the first columns of Tables \ref{tab:gApredLECs} and \ref{tab:gAfitLECs}, the $d_{16}$ value is slightly above the phenomenological one extracted in Ref.~\cite{Siemens:2017opr} and with much smaller error. Instead, $\g$ is in tension with the value obtained from experimental input, Eq. \eqref{eq:gAphys}. As a consequence $g_A (M_{\pi\rm (phys)}) = 1.205 \pm 0.010$, well below the experimental result. 
However, an inspection of the results for the $\pfour$ model described below reveals a large contribution from the $\pfour$ terms at $\mpi \gtrsim 200$ MeV, in line with the findings of Sec.~\ref{sec:gApred}. The fact that these terms are considerably larger than the error band of the $\pthree$ result implies that, in this case, the theoretical error estimated from $\mO(p^1)$ and $\pthree$  terms falls short in accounting for the $\pfour$ contribution. The agreement of  $g_A (\mpi)$ at $\pthree$ with LQCD points appears then as misleadingly good, while the uncertainties in the LEC values can be regarded as underestimated. 
In other words, to be realistic, an $\pthree$ analysis should be limited to small $\mpi < 200$ MeV, which would be unfeasible with the small amount of presently available LQCD data in this region.

\subsubsection{$\pfour$}

In the $\pfour$ $\slashed{\Delta}$ model one has the additional contribution of NLO LECs $c_{1-4}$. They were initially allowed to evolve in the fits under the constrains set by their empirical determination~\cite{Siemens:2017opr}  (second column of Table~\ref{tab:gApredLECs}) implemented as Gaussian priors. However, as these LECs are quite well determined, this procedure yields substantially the figures favored by the priors. Therefore, we report here the results of a simpler fit with $c_{1-4}$ held fixed to their central phenomenological values (second column of Table~\ref{tab:gAfitLECs}). 

\begin{figure}[h!]
\centering
\includegraphics[scale=0.85]{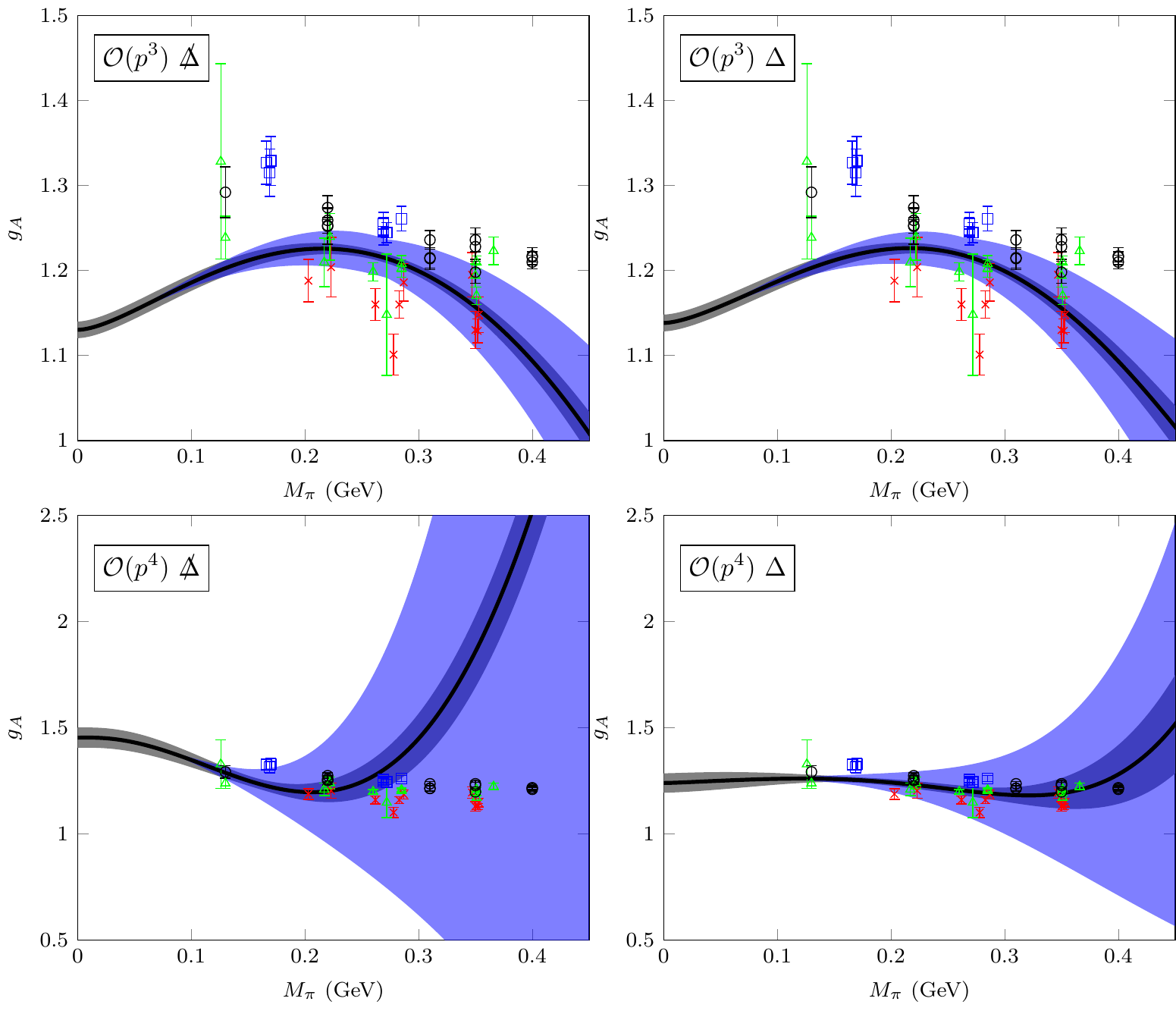}
\caption{Pion-mass dependence of $g_A$ obtained by fitting LQCD results with $\pthree$ and $\pfour$ relativistic BChPT without and with $\Delta(1232)$ as explicit degree of freedom. Gray (dark) bands correspond to errors determined by propagating LEC uncertainties.
Blue bands represent the estimated theoretical uncertainties $\Delta g^{(4,5)}_{A\chi}$. 
The LQCD points from CalLat~18~\cite{Chang:2018uxx} (black circles), Mainz~19~\cite{Harris:2019bih} (red crosses), RQCD~19~\cite{RQCD:2019jai} (green triangles) and NME~21~\cite{Park:2021ypf} (blue squares) are shown at their finite $a$ values i.e. without (small) discretization corrections.}
\label{fig:gAfit}
\end{figure}

In any case, as apparent from the lower left panel of Fig.~\ref{fig:gAfit} the $\pfour$ $\slashed{\Delta}$ model fails to describe the light-quark mass dependence of $g_A$.\footnote{Lattice-spacing corrections were neglected in this fit once they become unnaturally large (and uncertain) leading to overfitting.} The small $\chi^2$ is merely a consequence of the large theoretical error, which reduces the impact of high $\mpi$ points on the fit. Nevertheless, the poor agreement is reflected in the magnitude of $\chi_0^2$ and also in the quite unnatural  $d_{16}$ in spite of its prior. This large (in absolute value) $d_{16}$ is inconsistent with its  phenomenological value and nonetheless unable to correct the $\mpi$ dependence at $\mpi \gtrsim 300$~MeV, which is largely dominated by the $\pfour$ terms and, therefore, very similar to the one displayed in Fig.~\ref{fig:gApred}. 

The fact that the very wide theoretical error band encompasses the lattice points reflects that the disagreement would be removed by a (large) contribution of $\mO (p^5)$ counterbalancing the $\pfour$ ones. Actually, within heavy-baryon ChPT it has been shown~\cite{Bernard:2006te} that the curve can be {\it bent down} by additional contributions of orders $\mO (p^{5,6})$ with LECs of natural size (see also Fig.~1 of Ref.~\cite{Berengut:2013nh}). Here we take a different avenue and introduce the $\Delta(1232)$ explicitly as advocated in Ref.~\cite{Procura:2006gq} based on the Adler-Weisberger sum rule~\cite{Adler:1965ka,Weisberger:1965hp} and a heavy-baryon ChPT calculation for $g_A$.


\begin{table}[h]
\caption{LEC values, both fixed and fitted to LQCD data, in the four different models under study. The $\chi^2/\rm{dof}$ for each fit is given at the bottom; $\chi_0^2$, defined in Eq.~\eqref{eq:chi20}, does not include theoretical errors or naturalness priors.}
\centering
  \begin{tabular}{|c|c|c|c|c|}
  \hline
      & $\pthree$ $\slashed{\Delta}$ & $\pfour$ $\slashed{\Delta}$ & $\pthree$ $\Delta$  & $\pfour$ $\Delta$\\
    \hline
    $\g$ (free) & $1.1302\pm 0.0098$ & $1.453\pm 0.048$ & $1.1383\pm 0.0099$ &  $1.240\pm 0.046$ \\
    $d_{16}$ (GeV$^{-2}$) (free) & $-0.925\pm 0.055$ & $-9.77\pm 0.87$ & $1.224\pm 0.040$ & $-0.88\pm 0.88$ \\
    $h_A$    & -  & - & $1.35$ & $1.35$ \\
    $g_1$    & - & - & $-2.29$ & $-2.29$ \\
    $c_1$ (GeV$^{-1}$)  & -& $-0.89\pm 0.06$ & -  &  $-1.15\pm 0.05$ \\
    $c_2$ (GeV$^{-1}$)  & -& $3.38\pm 0.15$ & -   &  $1.57\pm 0.10$ \\
    $c_3$ (GeV$^{-1}$)  & -& $-4.59\pm 0.09$ & -  &   $-2.54\pm 0.05$ \\
    $c_4$ (GeV$^{-1}$)  & - & $3.31\pm 0.13$ & - & $2.61\pm 0.10$  \\
    $a_1$ (GeV$^{-1}$)  & - & -  & -  & $0.90$ \\
    $\widetilde{b}_4$ (GeV$^{-2}$) (free)  & - & - & - & $-12.3\pm 1.0$ \\
    $x_{1}$ (fm$^{-1}$)   & $0.39\pm 0.68$  & - & $0.38\pm 0.07$  &  $0.21\pm 0.07$ \\
    $x_{2}$ (fm$^{-2}$)    &  $-8.10 \pm 1.80$ & - &  $-8.17\pm 1.80$ &  $-7.27\pm 1.80$ \\
    $x_{3}$ (fm$^{-2}$)     & $2.25\pm 1.83$  & - & $2.17\pm 1.83$ &  $3.28\pm 1.84$ \\
    $x_{4}$ (fm$^{-1}$)     &  $0.61\pm 0.11$ & - &  $0.60\pm 0.11$ &  $0.51\pm 0.11$ \\
    $\m$ (GeV)    & 0.874 & 0.874  & 0.855  &  0.855\\
    $\md$ (GeV)    & - & - & 1.166 & 1.166 \\
    \hline
    $\chi^2/\rm{dof}$    & $36.06/(43-6)=0.98$ & $13.31/(43-2)=0.33$ & $37.60/(43-6)=1.02$ & $11.14/(43-7)=0.31$ \\
    $\chi_0^2/\rm{dof}$    & $424.87/(43-6)=11.48$   & $122820.67/(43-2)=2995.63$ & $439.19/(43-6)=11.87$  & $501.62/(43-7)=13.93$  \\
    \hline
  \end{tabular}
  \label{tab:gAfitLECs}
\end{table}

\subsection{\label{subsec:withD} $\Delta$ case}
With the $\Delta(1232)$ as explicit degree of freedom, new contributions with additional LECs arise. We fix the  $\mL_{\pi N\Delta}^{(1)}$ coupling $h_A$ to its large-$N_c$ value $h_A=1.35$~\cite{Siemens:2017opr}, which is close to its empirical value~\cite{Yao:2016vbz}. For the $\mL_{\pi\Delta}^{(1)}$ coupling $g_1$, whose impact on $g_A$ is small, the large-$N_c$ limit yields $|g_1| = 2.29$~\cite{Siemens:2017opr,Yao:2016vbz}. We are unable to discern its sign in our fits to $g_A$ LQCD data and opt for $g_1 = - 2.29$, preferred both by $\pi N$ elastic scattering~\cite{Yao:2016vbz} and our own studies of the nucleon axial form factor at low $q^2$ to be reported elsewhere. Finally, in analogy to the nucleon case, $\md \simeq m_{\Delta \rm (phys)} - 4 \, a_1 M_{\rm (phys)}^2$ with $m_{\Delta \rm (phys)}=1232$~MeV and $a_1 = 0.90$~GeV$^{-1}$ from the LQCD $\mpi$ dependence of the $\Delta$ mass~\cite{Alvarez-Ruso:2013fza}.   


\subsubsection{$\pthree$}

With this model, the result of the fit for $g_A(\mpi)$ closely resembles the $\pthree$ $\slashed{\Delta}$ one as can be seen in the comparison of the upper panels of Fig.~\ref{fig:gApred}. However, the value of $d_{16}$ changes considerably, including its sign, with respect to the $\pthree$ $\slashed{\Delta}$ one. The same feature was obtained when this LEC was extracted from LQCD results for the axial form factor at low $q^2$ with $\pthree$ relativistic BChPT~\cite{Yao:2017fym}. The $\pfour$ fit described in the next section produces $\pfour$ terms larger than the  theoretical error estimated from $\mO(p^1)$ and $\pthree$ ones (although to a lesser extent than in the $\slashed{\Delta}$ case) making the extension to $\pfour$ desirable for a more realistic determination of LECs and their uncertainties.


\subsubsection{\label{subsubsec:p4Dfull}$\pfour$}

In the fit of the $\pfour$ model with explicit $\Delta$ we fix the $c_{1-4}$ LECs to the values extracted from the $\pi N$ scattering~\cite{Siemens:2016hdi} (see the last column of Table~\ref{tab:gAfitLECs}). They take the $\Delta$ pole into account and are in good agreement with the joint $\pi N+\pi\pi N$ fit from Ref.~\cite{Siemens:2017opr}. In addition to $\g$ and $d_{16}$ we now have $\mO(p^2)$ $b_4$ and $b_5$ LECs as free parameters. As mentioned in Sec.~\ref{subsec:gATheo}, these LECs appear at $\pfour$ [more precisely, at $\mO(\mpi^4/\delta)$] in the combination $\bfour=b_4+ (12/13)\, b_5$. Therefore we keep only  $\bfour$ as a free parameter of the fit and neglect remaining higher order monomials proportional to $b_5$. 
\par 
The result of the fit, depicted in the lower right panel of Fig.~\ref{fig:gAfit}, satisfactorily describes the trend of $g_A(\mpi)$ as predicted by LQCD up to relatively large $\mpi$. The theoretical error is large and rapidly increasing with $\mpi$ due to the large contribution of $\pfour$ terms. Nevertheless, the description of LQCD data and convergence are notably improved with respect to the $\pfour$ $\slashed{\Delta}$ model. 

The extracted LECs are given in Table~\ref{tab:gAfitLECs}. The $\bfour$ value might seem unnatural but one should keep in mind that it is a combination of LECs. As shown in Table~\ref{tab:correlations} the correlations among LECs are sizable; they are an indication of degeneracy among the parameters that could be partially lifted by adding a new dimension to the fit (i.e. studying the axial form factor at finite $q^2$). 
\begin{table}[h!]
\caption{Correlation matrix for the fit with the $\pfour$ $\Delta$ model.}
\centering
  \begin{tabular}{|c|c c c|}
  \hline
      & $\g$ & $d_{16}$ & $\bfour$ \\
    \hline
    $\g$   & 1 & -0.97 & 0.79\\
    $d_{16}$   &  & 1 & -0.92 \\
    $\bfour$   &  &  & 1 \\
    \hline
  \end{tabular}
  \label{tab:correlations}
\end{table}
The $d_{16}$ value obtained from this model, $d_{16}= -0.88\pm 0.88$~GeV$^{-2}$, is in good agreement with the determination from inelastic $\pi N \raw \pi\pi N$  with explicit $\Delta$ pole~\cite{Siemens:2017opr} which, translated to standard EOMS, is $d_{16 \rm{(pheno)}}=-1.0\pm 1.0$~GeV$^{-2}$. Although, as argued in the Introduction, the $\mpi$ dependence of $g_A$ is in principle better suited to extract $d_{16}$ than the available experimental $\pi N \raw \pi\pi N$ data, the convergence issues of the former lead to large errors, comparable to those of the phenomenological result. The $\g$ value is higher than in the $\pthree$ fits and with a larger error. Furthermore, at the physical point $g_A (M_{\pi\rm (phys)}) = 1.260 \pm 0.012$ is now close to the experimental value although with a much larger uncertainty. 

\begin{figure}[h!]
\centering
\includegraphics[scale=0.85]{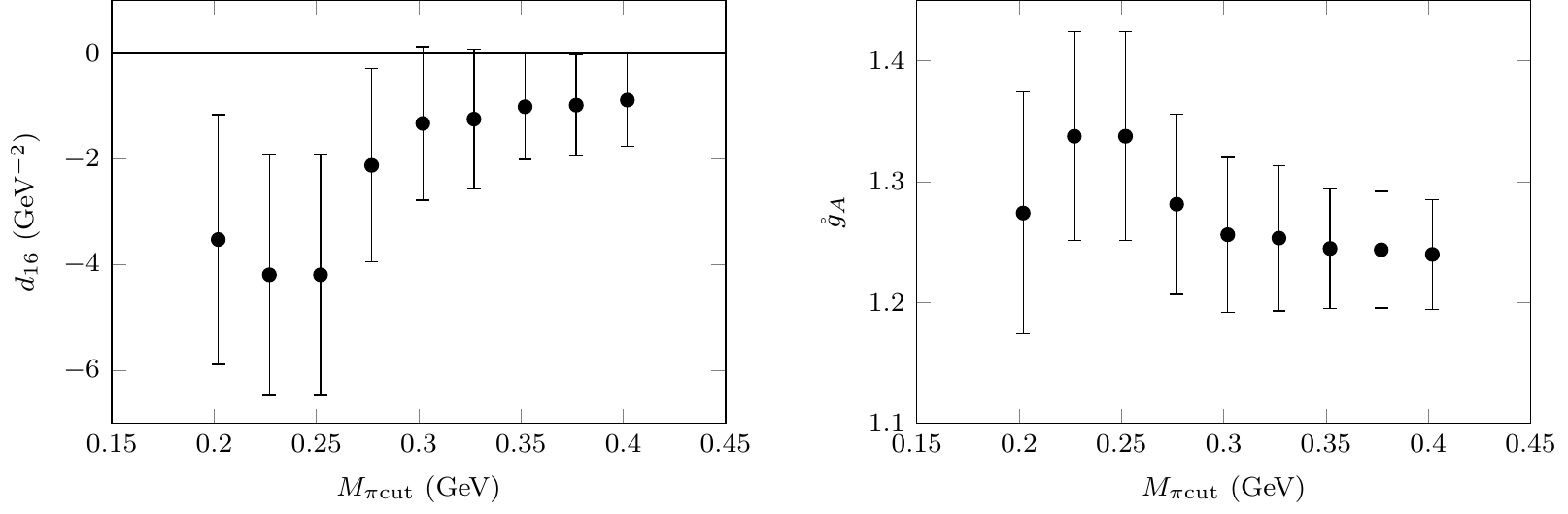}
\caption{Results of the fits for the $\pfour$ model with explicit $\Delta$ as a function of the maximum $M_{\pi \rm cut}$.}
\label{fig:mpicut}
\end{figure}

The stability of the results for $\g$ and $d_{16}$ as a function of the maximum $\mpi$ considered, $M_{\pi \rm cut}$, is shown in Fig.~\ref{fig:mpicut}. One can see that for both quantities the numerical values and their errors stabilize for $\mpi \gtrsim 300$~MeV.

Owing to the consideration of theoretical errors, the analysis could be extended to a broader range of $\mpi$ because points at higher $\mpi$, where the convergence is worse, have a larger combined uncertainty, Eq.~\eqref{eq:toterror}, and contribute less to the fit.

\section{\label{sec:concl} Conclusions and outlook}
We have investigated the pion mass dependence of the nucleon axial coupling up to $\pfour \equiv \mO(\mpi^3)$ (NNLO) in relativistic BChPT with EOMS renormalization. We have shown that, at this order and without including  $\Delta(1232)$ explicitly, the pion mass dependence of $g_A$ obtained using LECs extracted from phenomenological analyses of pion-nucleon elastic and inelastic ($\pi N \raw \pi \pi N$) scattering cannot describe the rather flat behavior predicted by state-of-the-art LQCD simulations. The disagreement is manifest from pion masses right above the physical point.
This feature, earlier observed in non-relativistic heavy-baryon ChPT~\cite{Bernard:2006te,Procura:2006gq}, is therefore also present in the relativistic theory. The fact that $\pfour$ terms become large from relatively low $\mpi \gtrsim 200$~MeV implies that $\pthree$ analyses of $g_A (\mpi)$ likely underestimate theoretical uncertainties.   

In line with the findings of Ref.~\cite{Procura:2006gq} within heavy-baryon ChPT,  we can satisfactorily describe the LQCD data for $g_A (\mpi)$ at $\pfour$ only after the $\Delta$ is incorporated as an explicit degree of freedom. However, although in a much smaller degree than in the $\slashed{\Delta}$ case, a  fast increase in the relative size of $\pfour$ terms with $\mpi$ is observed and reflected by the estimate of theoretical uncertainties. This fact jeopardizes the precision of the ChPT description of $g_A (\mpi)$ at $\mpi \gtrsim 300$~MeV and negatively influences the extraction of LECs based on LQCD data at such $\mpi$. Together with the sizable correlations, which can be reduced by considering both the $\mpi$ and $q^2$ dependence of the axial form factor in the fits, this feature implies that heavier resonances and/or $\mO(p^5)$ terms are still required to reach a good convergence and reduce theoretical errors. The impact of setting the baryon masses in the loops to the values obtained by the LQCD simulations may be also worth exploring in view of the results of Ref.~\cite{Lutz:2020dfi} although this would correspond to the resummation of baryon selfenergy insertions of higher order. For this purpose it would be relevant to have more information about the $\Delta$ pole position for the different lattice ensembles.    

From our $\pfour$ analysis of recent LQCD data we have obtained $g_A (M_{\pi\rm (phys)}) = 1.260 \pm 0.012$, close to the experimental value, and $d_{16}= -0.88\pm 0.88$~GeV$^{-2}$, in agreement with $\pi N$ phenomenology. As a consequence of the previously discussed issues, errors are still large, particularly for $d_{16}$ that is naturally extracted from the light-quark mass dependence of the nucleon axial coupling. New LQCD results at $\mpi \lesssim 250$~MeV will improve the precision of the extracted LECs. Besides, our ongoing effort to extend the analysis to the whole axial form factor (at low $q^2)$ may shed more light on $d_{16}$, as well as other LECs such as $d_{22}$.

\begin{acknowledgments}
We thank G. Guerrero, S. Leupold and M. J. Vicente Vacas for useful discussions, and D. Yao for sharing details of his related calculation with us. E. Epelbaum, A. Gasparyan and H. Krebs kindly answered our questions about their renormalization prescription. We are also indebted to the RQCD team for making their axial form-factor results available to us and to R. Gupta for providing relevant information about the NME ones.   
This research has been partially supported by the Spanish Ministerio de Ciencia e Innovaci\'on under contracts FIS2017-84038-C2-1-P and PID2020-112777GB-I00, the EU STRONG-2020 project under the program H2020-INFRAIA-2018-1, grant agreement no. 824093 and by Generalitat Valenciana under contract PROMETEO/2020/023.
\end{acknowledgments}

\appendix



\section{\label{app2} LEC conversion}

The covariant renormalization prescription adopted in Refs.~\cite{Siemens:2016hdi,Siemens:2017opr} differs from EOMS. In EOMS only power-counting breaking terms are absorbed in the LECs. To obtain an equivalence between covariant and heavy-baryon results, the prescription of Ref.~\cite{Siemens:2016hdi} also subtracts infrared-regular terms at the order of the calculation, as well as pieces proportional to $\log(\mpi^2/m_N^2)$. After setting $\bar{\lambda}=0$ and $\mu=m_N$ in Eq.~(21) of Ref.~\cite{Siemens:2016hdi} one is left with a dimensionally regularised LEC, $x_r$:
\begin{equation}
x_r = \bar{x}^{\rm{cov}} + \frac{1}{F_{\pi}^2} \left( \delta \bar{x}^{(3)}_f  + \delta \bar{x}^{(4)}_f \right) 
+ \frac{\beta_x}{32 \pi^2 F_{\pi}^2} \log{\left(\frac{M_{\pi\rm{(phys)}}^2}{m_{N\rm{(phys)}}^2} \right)} = x^{\rm{EOMS}} + \delta x_f^{\rm{EOMS}} \,
\end{equation}
which allows to express the renormalized EOMS LECs $x^{\rm{EOMS}}$ in terms of the corresponding ones, $\bar{x}^{\rm{cov}}$, from Refs.~\cite{Siemens:2016hdi,Siemens:2017opr}. The $\beta$-functions $\beta_{x}$ required to cancel the mesonic tadpole terms are listed in Appendix~B of Ref.~\cite{Siemens:2016hdi} as well as finite shifts $\delta \bar{x}^{(3,4)}_f$. Our EOMS finite shifts $\delta x_f^{\rm{EOMS}}$ are given in the supplementary material. In particular, for the relevant LECs in this study we obtain that 
\begin{equation}
\label{eq:d16conv}
d_{16} ^{\mathrm{EOMS}}=d^{\rm{cov}}_{16}+\frac{1}{F_{\pi}^2}\delta \bar{d}^{(3)}_{16f} + \frac{\beta_{d_{16}}}{32 \pi^2 F_{\pi}^2} \log{\left(\frac{M_{\pi\rm{(phys)}}^2}{m_{N\rm{(phys)}}^2} \right)} \,,
\end{equation}
with 
\begin{equation}
\label{eq:d16conv2}
\delta \bar{d}^{(3)}_{16f} = \frac{1}{32  \pi^2} ( \g+ \g^3 )    \,\,\, \mathrm{and} \, \,\, \beta_{d_{16}} = \frac{\g}{2} + \g^3 \,
\end{equation}
because, in this case, $\delta \bar{d}^{(4)}_{16f} = \delta d_{16f}^{\mathrm{EOMS}} $. Furthermore, 
\begin{equation}
\label{eq:cconv}
    c_{1-4 }^{\mathrm{EOMS}}=c_{1-4}^{\rm{cov}}
\end{equation}
once $\beta_{c_{1-4}} =0$~\cite{Siemens:2016hdi} and the finite shifts coincide with those in EOMS. Equations~\ref{eq:d16conv}-\ref{eq:cconv} are derived for the $\slashed{\Delta}$ case but hold also for the model with the $\Delta$ pole of Ref.~\cite{Siemens:2016hdi} as it does not involve additional renormalizations.



\section{\label{app3} Contribution to $g_A$ from wave function renormalization}


To the order of the present study $Z_N$, defined in Eq.~\ref{eq:wfr1}, can be written as $Z_N=1+\frac{1}{16\pi^2\fpi^2 }\left(\delta Z_N ^{(3)\rm{loop}}+\delta Z_N ^{(4)\rm{loop}}\right)$. The contribution to $g_A$ that starts at $\pfour$ is then $\g \delta Z_N ^{(4)\rm{loop}}$, with $\delta Z_N ^{(4)\rm{loop}}$, including $\Delta$, given by
\begin{widetext}
\bea
    & &\delta Z_N ^{(4)\rm{loop}}=-\frac{24  c_1 \g^2  \mpi^5 f(\mpi)}{\m^3 a^{3/2}} \left\{6 \m^4-6 \mpi^2 \m^2+\mpi^4 \right\}\no\\
    &+& \frac{2 \mpi^4}{27 \m^3} \left\{\frac{a_1 h_A^2 \m}{\md^4} \left(18 \m^4 \md-4 \m^3 \left(\mpi^2-9 \md^2\right)+\m^2 \left(36 \md^3-103 \mpi^2 \md\right)-36 \m \md^2 \left(\mpi^2+4 \md^2\right)\right.\right.\no\\
    &+&\left.\left. 54 \md \left(\mpi^4+\mpi^2 \md^2-5 \md^4\right)\right) +\frac{18 c_1}{\md^2} \bigg(\frac{18 \g^2 \m^2 \md^2 \left(\mpi^2-2 \m^2\right)}{a}+h_A^2 \left(8 \m^4+6 \m^3 \md \right.\right.\no\\
    &+&\left.\left.\m^2 \left(\mpi^2-2 \md^2\right)6 \m \md \left(\mpi^2-2 \md^2\right)-6 \left(\mpi^4-3 \mpi^2 \md^2+3 \md^4\right)\right)\bigg)\right\}\no\\
    &+& \frac{4 \mpi^4}{9 \m^3\md^4}  \log{\left(\frac{\m^2}{\mpi^2}\right)} \bigg\{a_1 h_A^2 \m \left(-9 \m^4 \md+\m^3 \left(4 \mpi^2-6 \md^2\right)+13 \m^2 \mpi^2 \md-3 \mpi^4 \md+3 \md^5\right)\no\\
    &-& 27  c_1 g^2 \md^4 \mpi^2\bigg\}\no\\
    &+&\frac{8}{3} h_A^2 \mpi^2  \log{( \m )} \left\{a_1 \left(\frac{2 \md^3}{\m^2}-2 \m+\md\right)-\frac{6 c_1 \mpi^2 (2 \m+\md)}{\md^2}\right\}\no\\
    &-& \frac{8 c_1 h_A^2 \mpi^2  \log{( \mpi )}}{3 \m^5 \md^2}  \Big\{5 \m^8+6 \m^7 \md-3 \m^6 \left(2 \mpi^2+\md^2\right)-3 \m^5 \md \left(\mpi^2+\md^2\right)   \no\\
    &+& \m^2 \left(2 \mpi^6-3 \mpi^4 \md^2+\md^6\right)+3 \m \md \left(\mpi^2-\md^2\right)^3-3 \left(\mpi^2-\md^2\right)^4\Big\}\no\\
    &-& \frac{4 a_1 h_A^2 \mpi^2}{ 3 \m^2 \md^3}  \log{\left(\frac{\m^2}{\md^2}\right)} \Big\{9 \m^4 \mpi^2+\m^3 \left(4 \mpi^2 \md-2 \md^3\right)+\m^2 \left(\md^4-3 \mpi^4\right)-\mpi^6+\mpi^2 \md^4+2 \md^6\Big\}\no\\
    &+& \frac{2 a_1 h_A^2 \mpi^2 }{3 \m^4 \md^3} \log{\left(\frac{\mpi^2}{\md^2} \right)} \Big\{-5 \m^8-4 \m^7 \md+6 \m^6 \mpi^2+2 \m^5 \md \left(\mpi^2-3 \md^2\right)\no\\
    &+&2 \m^2 \left(\mpi^6+\mpi^2 \md^4-2 \md^6\right)+2 \m \md \left(\mpi^2-\md^2\right)^2 \left(\mpi^2+5 \md^2\right)-3 \left(\mpi^2-\md^2\right)^3 \left(\mpi^2+3 \md^2\right)\Big\}\no\\
    &-& \frac{8 h_A^2 \mpi^2}{3 \m^5 \md^3} \log{(\md)}\Big\{a_1 \m \left(5 \m^8+4 \m^7 \md+4 \m^5 \md^3+\m^4 \md^4+6 \m^2 \md^6-10 \m \md^7-9 \md^8\right)\no\\
    &-& c_1 \md \left(-5 \m^8-6 \m^7 \md+3 \m^6 \left(2 \mpi^2+\md^2\right)+3 \m^5 \md \left(\mpi^2+\md^2\right)+\m^2 \left(2 \mpi^6-3 \mpi^4 \md^2-\md^6\right)\right.\no\\
    &+&\left.3 \m \md \left(\mpi^6-3 \mpi^4 \md^2+3 \mpi^2 \md^4+\md^6\right)-3 \left(\mpi^8-4 \mpi^6 \md^2+6 \mpi^4 \md^4-4 \mpi^2 \md^6-\md^8\right)\right)\Big\}\no\\
    &+& \frac{4 h_A^2 \mpi^2 (\m-\md) (\m+\md)^3}{3 \m^5 \md^3} \log  \left(\md^2-\m^2\right) \Big\{a_1 \m \left(5 \m^4-6 \m^3 \md+12 \m^2 \md^2-8 \m \md^3+9 \md^4\right)\no\\
    &+& 2 c_1 \md \left(5 \m^4-4 \m^3 \md+5 \m^2 \md^2-3 \m \md^3+3 \md^4\right)\Big\}\no\\
    &+& \frac{F( \mpi )}{3 \m^5 \md^3}\Bigg\{4 h_A^2 \left(\m^2+2 \m \md-\mpi^2+\md^2\right) \left(a_1 \m \bigg(5 \m^8-6 \m^7 \md+2 \m^6 \left(\md^2-6 \mpi^2\right)\right.\no\\
    &+&\left. 4 \m^5 \left(2 \mpi^2 \md+\md^3\right)+2 \m^4 \left(3 \mpi^4-5 \md^4\right)+2 \m^3 \md \left(\mpi^4-6 \mpi^2 \md^2+5 \md^4\right)\right.\no\\
    &+& \left. \m^2 \left(4 \mpi^6-2 \mpi^4 \md^2+4 \mpi^2 \md^4-6 \md^6\right)-4 \m \left(\mpi^6 \md-3 \mpi^2 \md^5+2 \md^7\right)\right.\no\\
    &-& \left.3 \left(\mpi^2-\md^2\right)^3 \left(\mpi^2+3 \md^2\right)\bigg)\right.\no\\
    &+& \left. 2 c_1 \md \bigg(5 \m^8-4 \m^7 \md-\m^6 \left(6 \mpi^2+5 \md^2\right)+\m^5 \left(5 \md^3-\mpi^2 \md\right)+2 \m^4 \md^2 \left(\mpi^2-\md^2\right)\right.\no\\
    &+& \left. 2 \m^3 \md \left(\mpi^2-\md^2\right)^2-\m^2 \left(2 \mpi^6-3 \mpi^4 \md^2+\md^6\right)+3 \m \md \left(\mpi^2-\md^2\right)^3+3 \left(\mpi^2-\md^2\right)^4\bigg)\right)\Bigg\},
\eea
\end{widetext}
with the following definitions:
\begin{equation}
    F(\mpi)\equiv\frac{\mpi^2}{b}\log{\left( \frac{-\m^2+b+\mpi^2+\md^2}{2 \mpi \md}\right)},\quad f(\mpi) \equiv i \log{\left(\frac{\mpi+i \sqrt{a}}{2\m}\right)},
\end{equation}
where $b\equiv\sqrt{\m^4-2 \m^2 \left(\mpi^2+\md^2\right)+\left(\mpi^2-\md^2\right)^2}$ and  $a\equiv4\m^2-\mpi^2$. This result, but also the full expression for the $\pfour$ nucleon selfenergy $\Sigma_N$ with explicit $\Delta$, from which $Z_N$ is derived using Eq.~\ref{eq:wfr1}, can be found in the Mathematica notebook provided as supplementary material.


\bibliography{axialcharge}

\end{document}